\documentclass[showpacs, amsmath, amssymb, amsthm, aps, jmp, thmsb, twocolumn, superscriptaddress]{revtex4-2}

\usepackage[dvipdfmx]{graphicx}
\usepackage{color}
\usepackage{float}
\usepackage{amsmath}
\usepackage{amssymb}
\usepackage{amsfonts}
\usepackage{amsbsy}
\usepackage{latexsym}
\usepackage{mathrsfs}
\usepackage{bbm}
\usepackage{bm}

\newtheorem{lemma}{Lemma}

\newcommand{\be}{\begin{eqnarray}}
\newcommand{\ee}{\end{eqnarray}}
\def\({\left(}
\def\){\right)}
\def\[{\left[}
\def\]{\right]}
\def\C{\mathbb{C}}

\newcommand{\bra}[1]{\langle #1 |}
\newcommand{\ket}[1]{| #1 \rangle}

\newcommand{\sla}[1]{\rlap{\kern .15em /}#1}

\newcommand{\Bot}{\bigotimes}

\newcommand{\tr}{{\rm tr}}

\begin{document}
\title{Bayesianism, Conditional Probability and Laplace Law of Succession\\ in Quantum Mechanics}

\author{Tsubasa Ichikawa}
\affiliation{Center for Quantum Information and Quantum Biology, Osaka University, Osaka 560-0043, Japan.}
\email{Present address: Department of Physics, The University of Tokyo, Bunkyo-ku, Tokyo 113-0033, Japan.\\
email: tsubasa.ichikawa@phys.s.u-tokyo.ac.jp}

\begin{abstract}
 We present a comparative study between classical probability and quantum probability from the Bayesian viewpoint, where probability is construed as our rational degree of belief on whether a given statement is true. 
 From this viewpoint, including conditional probability, three issues are discussed:
 i) Given a measure of the rational degree of belief, does it satisfy the axioms of the probability?
 ii) Given the probability satisfying these axioms, is it seen as the measure of the rational degree of belief?
 iii) Can the measure of the rational degree of belief be evaluated in terms of the relative frequency of events occurring?
Here we show that as with the classical probability, all these issues can be resolved affirmatively in the quantum probability, provided that the relation to the relative frequency is slightly modified from the Laplace law of succession in case of a small number of observations. 
This implies that the relation between the Bayesian probability and the relative frequency in quantum mechanics is the same as that in the classical probability theory, including conditional probability.
\end{abstract}

\maketitle

\section{Introduction}

Probability theory allows two mutually opposite viewpoints on its interpretation \cite{Laplace02, Gillies00}.
One is the Bayesian viewpoint \cite{Kaynes21,deFinetti37,deFinetti92,Cox1946_nature,Cox46,Cox61,Shimony55,Kemeny55, Lehman55, Jaynes1989, Jaynes1990, Jaynes03, vanHorn03}, where the probability is understood as our degree of belief on how a given statement is true under incomplete information. In other words, probability is an epistemic concept, measuring how our subjective knowledge is likely.
The other is the ontological viewpoint, where probability represents a definite objective property, and it can be measured by the relative frequency of the repeatable observations.

An attempt to a reconciliation of these two viewpoints was put forward by de Finetti \cite{deFinetti37,deFinetti92} and extended in \cite{Shimony55, Kemeny55, Lehman55}.
The strategy is as follows:
i) Assuming that the degree of belief is measured by bets, and defining a condition called coherence, which ensures rationality of one\rq s degree of belief.
ii) Proving that if the degree of belief satisfies the coherence condition \cite{deFinetti37,deFinetti92, Shimony55}, then it satisfies the axioms of probability theory.
iii) Proving the converse of ii), that is, showing that if the degree of belief satisfies the axiom of probability, then it fulfills the coherence condition \cite{Kemeny55, Lehman55}.
iv) Demonstrating that the degree of belief approaches the relative frequency for infinitely repeatable and exchangeable events for which we have no prior information.
Steps ii) and iii), called Dutch Book Argument (DBA) and the converse DBA, respectively, ensure the equivalence between the rational degree of belief and probability.

In the last two decades, there is much literature to derive quantum mechanical probability (the Born rule) from the Bayesian viewpoint \cite{Fuchs13}.
Among them, one of the influential works is \cite{Caves02}, which derived the Born rule by resorting to the quantum analogue of the DBA and discussed the relation to the relative frequency by using a simple combinatorial argument to independent and identically distributed (i.i.d.) arrays of the pure states.
In comparison with de Finetti\rq s strategy for the reconciliation, the following three analyses are still requisites to extend the scope of the application of the Bayesian viewpoint in quantum mechanics (QM): the quantum analogue of the DBA including the conditional probability, the converse DBA, and the argument about the relative frequency for the infinitely repeatable and exchangeable events on the quantum systems.
Here and hereafter we use the term QM to refer its probabilistic aspects described by the Hilbert space formalism, not the physical phenomena described by QM.

This paper performs a comparative study between probability theory and QM from the Bayesian viewpoint.
To achieve this, we perform the mandatory analyses mentioned in the previous paragraph and shows a coherent picture of the Bayesian viewpoint and relative frequency in QM.
More precisely, we extend the quantum analogue of the DBA utilized in \cite{Caves02} in such a way to include conditional probability and consider its converse. 
Through these considerations, we show that our degree of belief satisfying the coherence condition takes the form of the Born rule and the quantum conditional probability associated with the L\"uders projection postulate \cite{Luders50, Busch09}, and the converse DBA holds even when the non-commutativity of projectors are taken into account.
These suggest that, similarly to probability theory, quantum probability is fully understandable from the Bayesian point of view.

Moreover, our analyses uncover the relation between quantum probability and relative frequency.
In the Bayesian viewpoint, the (classical) probability approaches the relative frequency through the Laplace law of succession \cite{Laplace02} under no prior information.
We will show that this is true even in QM, with the slight modification to the Laplace law of succession.
Thus, summing up, the present work makes the counterpart in QM to the classical argument for the Bayesian probability and relative frequency.

The rest of the paper is organized as follows:
In Sec.~\ref{OR}, we summarize our motivation, research questions, and the results to be obtained in this paper.
In Sec.~\ref{c_dba}, we review three issues raised in the Bayesian viewpoint in classical probability, that is, the DBA, the converse of the DBA, and the relation between the probability and relative frequency.
In Sec.~\ref{q_dba}, we extend the DBA and its converse to QM.
We derive the Born rule and the concrete expression of quantum conditional probability by using the DBA, and give a proof of the converse of the DBA.
In Sec.~\ref{q_rel_freq}, we show that quantum probability approaches the relative frequency via the modified Laplace law of succession, by using the quantum de Finetti representation theorem \cite{Caves02b}.
Section~\ref{conc} is devoted to our conclusion and discussions.
From Sec.~\ref{c_dba} to Sec.~\ref{q_rel_freq}, we present the whole picture of the Bayesian constructions of probability theory and QM with technical details, and readers who are uninterested in the details may skip these sections.

\section{Backgrounds and results}
\label{OR}

As announced, this paper provides a comparative study between classical and quantum probability from the Bayesian viewpoint.
In this section, we summarize the necessary backgrounds, the research questions, and the results to be obtained in the following sections.

\subsection{Motivation}
\label{motiv}
Quantum mechanics is a theory describing the outcomes of measurements on microscopic systems.
Its descriptions of experiments presuppose the measured system (object) and the measurer (subject) \cite{Bell1973}.
Heisenberg was aware of the subject-object distinction in the standard interpretation of QM and its consequence:
\begin{quotation}
A real difficulty in the understanding of this [Copenghagen] interpretation arises, however, when one asks the
famous question: But what happens `really' in an atomic event? It has been said before that the mechanism and the results of an observation can always be stated in terms of the classical concepts. But what one deduces from an observation is a probability function, a mathematical expression that combines statements about possibilities or tendencies with statements about our knowledge of facts. So we cannot completely objectify the result of an observation, we cannot describe what `happens' between this observation and the next. \cite{Heisenberg1958-HEIPAP}
\end{quotation}
This statement suggests that QM is a formalism that inherently and implicitly contains information-acquiring processes, which are of the measurer\rq s activities.

The coexistence of subject and object is not problematic for all practical purposes.
As Heisenberg pointed out, however, it triggers conceptual difficulty if one wishes an ontological picture of the microscopic phenomena.
For the analysis of this difficulty, the resolution of the subjective and objective parts in QM would be  desirable, as stated by Jaynes:
\begin{quotation}
But our present QM formalism is not purely epistemological; it is a peculiar mixture describing in part realities of Nature, in part incomplete human information about Nature - all scrambled up by Heisenberg and Bohr into an omelette that nobody has seen how to unscramble. Yet we think that the unscrambling is a prerequisite for any further advance in basic physical theory. For, if we cannot separate the subjective and objective aspects of the formalism, we cannot know what we are talking about; it is just that simple. So we want to speculate on the proper tools to do this. \cite{Jaynes1990}
\end{quotation}

The tool Jaynes chose was the Bayesian probability, on the ground of resemblance between Bohr\rq s reply \cite{Bohr1935} to the Einstein-Podolsky-Rosen (EPR) argument \cite{EPR1935} and the Bayesian inference \cite{Jaynes1989}.
Here Bayesian probability means the schools of interpretations of probability, where the probability is understood as a measure of our degree of belief on how a given statement is true under incomplete information.

We hereafter employ the Bayesian probability, as Jaynes and QBists \cite{Caves02,Caves02b,Fuchs13,Fuchs2014,vonBaeyer2016} did so. 
In terms of the Bayesian probability, the resolution of the objective and subjective parts in QM could be reduced to answering the following question: what and how much of the formalism of QM is explained by the Bayesian probability? Something unexplainable in terms of the Bayesian probability, if exists, could be objective.
To find it, all the probabilistic aspects of QM should be investigated.
We accomplish this programme in this paper.
Note that this programme means the expansion of the scope of application of the Bayesian viewpoint in QM too.

In passing, we point out that the Bayesian probability has strong affinities to pragmatism, which is a broad spectrum of schools of philosophy aiming at  \lq\lq knowing the world as inseparable from agency within it\rq\rq~\cite{sep-pragmatism}. 
Indeed, it has been known that several founders of the Bayesian probability were influenced by pragmatism \cite{Galavotti2019}, and founders of QBism also express the influence, especially from the pragmatist James \cite{Fuchs13}.
Accordingly, the present work is, by construction, also under the strong influence of pragmatism, because we hereafter fill the gap between the Bayesian probability and QBism, both of which share the influence of pragmatism.


\subsection{Problem Setting}
\label{PS}

With the above motivation, we answer the questions by filling the gap between the present achievements of QBism and the corresponding works in classical probability theory given mainly by de Finetti.
We chose these two schools because QBism is progressing along the Jaynes\rq~programme with a more subjective approach, which is partially based on the de Finetti\rq s works.

To identify the gap, we first make a review of the de Finetti\rq s arguments in the classical probability theory.
For details, see Sec.~\ref{c_dba}.
We next state the outline of QBism associated with the de Finetti\rq s argument.
We shall compare these and find the gap.

In de Finetti\rq s argument, we first suppose that our degree of belief can be quantified using bets.
More precisely, given a bet on whether a proposition $a$ is true, the ratio between the bettor\rq s (maximal) wager and the stake that the bettor would win measures the bettor\rq s degree of belief on how the statement is certain.
We call this ratio the {\it betting quotient} with respect to the proposition $a$, hereafter denoted by $q(a)$, or $q$ for short.
On the assumption of measurability of the degree of belief, we further introduce a consistency condition, called {\it coherence}, to the bettor\rq s set of the betting quotients on several bets.
Here the coherence is the requirement that the set of the betting quotients should be constructed in such a way that the bettor wins at least one combination of the outcomes of the bets under consideration.
Note that the coherence condition is a necessary condition for the rationality of the bettor: without coherence, the bettor is allowed to make the bets in such a way that he never wins and must have a sure loss.
As such, we hereafter call the degree of belief satisfying the coherent condition the coherent degree of belief or the rational degree of belief.

\subsubsection{Dutch Book Argument.}
Despite its simplicity, the coherence condition derives all the axioms of the probability theory.
In other words, the set of the betting quotients satisfying the coherence condition fulfills the axiom of probability:
\begin{subequations}
\be
&&q\ge0, \label{positivity}\\
&&q(a\lor b)=q(a)+q(b)\quad {\rm if}\quad a\land b=\emptyset, \label{additivity}\\
&&q(\Omega)=1, \label{tautology}\\
&&q(a\land b)=q(a|b)q(b). \label{c_cond}
\ee
\label{prob_axioms}
\end{subequations}
The argument deriving Eqs.~(\ref{prob_axioms}) is called the DBA.
Here, $\Omega$ and $\emptyset$ are the always true proposition (tautology) and the always false proposition (contradiction), respectively. 
Together with the NOT ($\lnot$) operation, AND operation ($\land$) and OR operation ($\lor$), the set of all the propositions forms the Boolean algebra. 
The quantity $q(a|b)$ is the conditional betting quotient of the proposition $a$ given $b$, which will be defined in Sec.~\ref{c_dba} by introducing conditional bettings.

\subsubsection{Converse Dutch Book Argument.}
Moreover, we can strengthen the above observation by proving the converse of the DBA  \cite{Lehman55,Kemeny55}, that is, {\it any} set of the betting quotients is coherent if it satisfies Eqs.~(\ref{prob_axioms}).
Together with the DBA, the converse DBA shows that nothing is inconsistent to assume that the probability is exactly the same as the one\rq s {\it rational} degree of belief.
Since the rational degree of belief is subjective notion, this equivalence between probability and the rational degree of belief shows that we need not introduce objective probability in order to understand the probability theory.
Note that the converse DBA deals with any set of betting quotients including those for conditional bettings. Thus, the DBA with the conditional bettings and its coherence condition is a prerequisite to proving the converse DBA.

\subsubsection{Relative Frequency.}
Since the probability is found to be subjective in our approach, the probability assignment on a given event may depend on the one\rq s subjectivity, and thereby it may become arbitrary, as far as one\rq s assignment obeys the axioms of probability. 
In contrast, the relative frequency is uniquely determined within the error margin and thereby considered as an objective concept; and it has been indeed employed in actual experiments. 
Bearing these in mind, we may ask a further question of whether we can explain the relative frequency using the Bayesian notions.

For this question, one can make an argument by considering the case that there are independent repeatable events with constant but unknown probability $p$.
For example, consider a coin-flipping game, where the coin is biased and the probability of finding its head in a toss is given by an unknown constant probability $p$. 
Then the probability to find $k$ heads in the $n$ tosses is proportional to the binomial distribution $\tbinom{n}{k}p^k(1-p)^{n-k}$, whose maximum is attained at $p=k/n$ and the probability of finding the relative frequency $k/n$ close to $p$ approaches the unity in large $n$ limit \cite{Cox46}.

Although the above example properly derives the relative frequency, it is difficult to justify the assumption of the constantness of $p$ from the Bayesian viewpoint, since the probability assignment for $p$ may depend on one\rq s subjectivity, and thereby arbitrary as stated before \cite{deFinetti37,deFinetti92}.
Thus we need to derive the relative frequency under more relaxed conditions.

Now, instead of the independent repeatable events with constant but unknown probability, let us consider the infinitely exchangeable events for which we have no prior information.
We will give the precise definition of infinite exchangeability in Sec.~\ref{c_dba}.
As to be shown later, the infinite exchangeability assumption corresponds with introducing a probability distribution that describes the observer\rq s subjective likelihood on how much the coin is biased.
On this setup, given the set ${\cal M}_{n,k}$ of the outcomes that we find the $k$ heads in the $n$ tosses, the conditional probability $q(k_{n+1}|{\cal M}_{n,k})$ to find the head at the $(n+1)$-th toss is given as
\be
q(k_{n+1}|{\cal M}_{n,k})=\frac{k+1}{n+2},
\label{laplace}
\ee
which is called the Laplace law of succession \cite{Laplace02, Gillies00, deFinetti37, deFinetti92, Cox46, Cox61}, or Laplace smoothing in machine learning community \cite{manning08retrieval}.
Clearly, the Laplace law of succession shows that the conditional probability approaches the relative frequency in the large $n$ limit.

\subsubsection{Remaining issues as counterparts lacking in QBism.}
QBism derived the Born rule and argued the relative frequency  \cite{Caves02}, through the identification between the proposition $a$ in classical probability and the projector $P$ in QM.
The derivation of the Born rule using the DBA implies that the density operator describes a subjective state of knowledge of an observer, from which his rational degree of belief is evaluated.
On the basis of these observations, QBism provides subjective explanations of genuine quantum phenomena such as the violation of the Bell inequality \cite{Fuchs2014,vonBaeyer2016}.

The Born rule gives the concrete form of the probability with respect to the quantum systems, but it does not give  the concrete form of the conditional probability;
it is still unanswered whether the quantum conditional probability is explained with the Bayesian notions.
In comparison with the argument in probability theory, conditional probability in QM should be also essential to develop the converse DBA and the relative frequency derivation in QM. Moreover, the rigorous treatment of the two-slit experiment requires the use of conditional probability \cite{BC81}. 

The derivation of the relative frequency in \cite{Caves02} assumes that an observer of the repeatable events has the maximal knowledge of the quantum system.
This assumption allows the observer to uniquely assign a pure state to the system.
Performing single-system projective measurements on the i.i.d. array of the assigned pure state, we obtain the multinomial distribution of the measurement outcomes, whose maximal is attained when the probability of finding the measurement outcome is given by the relative frequency.

The point of the above argument is the use of the maximal knowledge on a quantum system, which ensures one\rq s unique pure state assignment and thereby the probability of the measurement outcomes is inevitably constant, unlike the classical probability.
However, putting this unavoidability aside, the above argument  is limited to the pure state cases and still formally corresponds to the classical case of infinitely repeatable events with constant unknown probability.
Thus, we may expect that the relative frequency in QM could be derived on the ground of more relaxed conditions.

Summing up, there are two remaining issues on the Bayesian explanation of QM:
One is the derivation of the concrete expression of the conditional probability using DBA, followed by the converse DBA.
The other is the derivation of the relative frequency not employing the assumption of the maximal knowledge of the quantum system.
We shall perform these analyses in Secs.~\ref{q_dba} and \ref{q_rel_freq}.

\subsection{Results}
\label{sum_res}

In this paper, we fill the gap between classical probability and QM from the Bayesian viewpoint by deriving three results, which are summarized as follows.
First, by extending the DBA to the conditional betting on the projector $P$ given the outcome of the measurement of a projector $Q$, we obtain the explicit expression of the quantum conditional probability
\be
q(P|Q)=\tr(\rho_QP),
\label{q_cond}
\ee
where
\be
\rho_Q=\frac{Q\rho Q}{\tr(\rho Q)}
\label{rhoq}
\ee
is the post-measurement state with respect to the projector $Q$ defined with an appropriate density operator $\rho$.
More precisely, if the conditional betting quotient is coherent, then it takes the form (\ref{q_cond}).

This derivation of the expression (\ref{q_cond}) implies that the density operators $\rho$ and $\rho_Q$ represent our subjective knowledge about the given quantum system, since $q(P|Q)$ measures our degree of belief for the projector $P$ conditioned by $Q$.
However, they may behave as objective quantities in the following two cases:
i) If $\rho$ is pure, then there exists a normalized state vector $|\psi\rangle$ such that $\rho=|\psi\rangle\langle\psi|$, and we may further set $P=|\psi\rangle\langle\psi|$ to obtain $P\rho=\rho$.
Therefore, the outcome of the measurement of the projector $P$ given the density operator $\rho$ is unity with certainty, and thereby objective \cite{CAVES2007255}.
ii) In parallel with case i), we have $Q\rho_Q=\rho_Q$, showing that the outcome of the measurement of the projector $Q$ given the density operator $\rho_Q$ is unity with certainty, and thereby objective.
In the Bayesian viewpoint, such apparent objectivity of quantum states can be understood in terms of degrees of belief by reading the statement that the measurement outcome is 1 with certainty as a proposition of a one\rq s belief on the system \cite{CAVES2007255}.

Equation (\ref{q_cond}) was already found in connection with the L\"uders projection postulate \cite{Luders50, Busch09}.
Our derivation gives an interesting interpretation of the L\"uders projection postulate from the Bayesian viewpoint: Observer\rq s state of knowledge $\rho$ changes into $\rho_Q$ as he acquires the knowledge of the measurement outcome with respect to $Q$.
Given the updated state of knowledge, one\rq s rational degree of belief is evaluated as the conditional probability for the given quantum system.
Although this interpretation itself is of long-standing \cite{vonBaeyer2016,vonNeumann2018}, we gave its new derivation in a way precisely consistent with the Bayesian viewpoint.

Moreover, we may further extend the above observation.
Let us consider $n$ conditional bets on the projector $P$ conditioned by $Q_i$, where $i=1,\dots,n$. 
Given the rational degrees of belief $q(P|Q_i)$ and $q(Q_i)$, we can average these degrees of belief as
\be
\langle q(P)\rangle=g_1q(P|Q_1)+\dots+g_nq(P|Q_n),
\ee
where
\be
g_i=\frac{q(Q_i)}{q(Q_1)+\dots+q(Q_n)}
\ee
is the relative weight of the degree of belief that the conditioning $Q_i$ occurs.
From the linearity of the trace, we may write
\be
\langle q(P)\rangle=\tr(\rho_{\langle Q \rangle}P)
\ee
with the use of the state of our knowledge averaged over the possible conditionings
\be
\rho_{\langle Q \rangle}&=&g_1\rho_{Q_1}+\dots+g_m\rho_{Q_m}=\frac{\sum_{i=1}^mQ_i\rho Q_i}{\sum_{i=1}^mq(Q_i)},
\label{ave_rhoq}
\ee
which reduces to the aggregated post-measurement state already postulated by L\"uders when the outcomes of the measurements of $\{Q_i\}_{i=1}^n$ are unknown \cite{Luders50}.
Thus, the coherent degree of belief can describe two post-measurement states: Eq.~(\ref{rhoq}) for the direct measurements, and Eq.~(\ref{ave_rhoq}) for the measurements with the outcome aggregation.

Second, armed with Eqs.~(\ref{q_cond}) and (\ref{rhoq}), we extend the converse DBA to quantum systems and show that if the (conditional) betting quotients follow the Born rule and Eq.~(\ref{q_cond}), then the set of the (conditional) betting quotients is coherent.
This is accomplished in a way parallel with the argument in classical probability.

Combining the DBA and converse DBA, we may conclude that even in QM, nothing is inconsistent to assume that the probability is equivalent to the rational degree of belief.
In other words, to understand quantum probability, we need not introduce the notion of objective probability such as the \lq\lq statements about possibilities or tendencies\rq\rq~of the system Heisenberg addressed in the quotation.
Note that for the derivation of the equivalence between quantum probability and the rational degree of belief, we need the converse DBA, which is first extended to quantum systems in this work.
Note also that since the (conditional) probability is evaluated from the density operator, the converse DBA ensures the consistency of the interpretation that the density operator is the state of knowledge. 


\begin{figure}[t]
\begin{center}
(a) Flat measure\\
\includegraphics[width=3.5in]{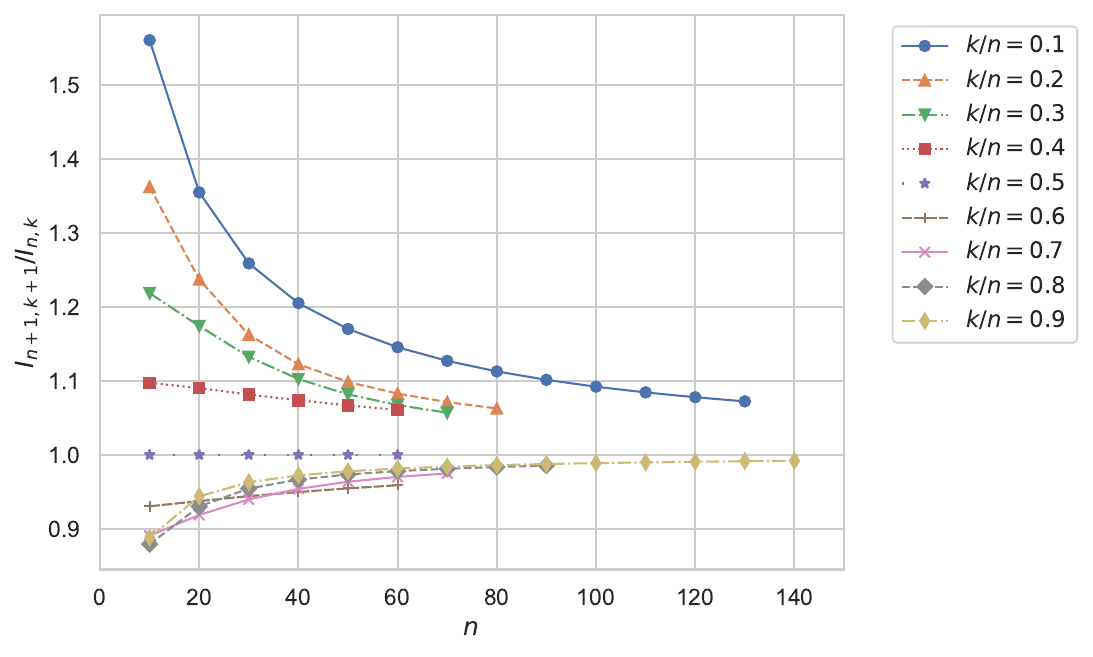}\\
(b) Bures measure\\
\includegraphics[width=3.5in]{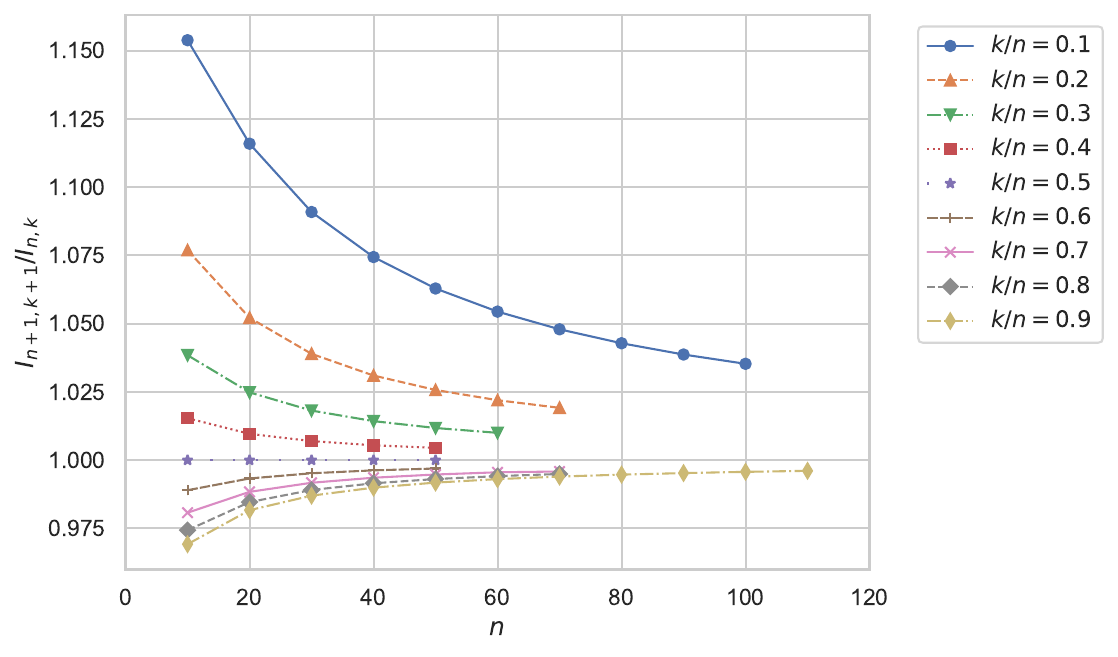}
\caption{(Color online) The correction term $I_{n+1,k+1}/I_{n,k}$ to the Laplace law of succession for (a) the flat measure and (b) the Bures measure. The correction term tends to unity for large $n$ in both cases, irrespective of the value of the relative frequency $k/n$. Note that we have not plotted the correction term  for the uniform measure of the pure state space, since $I_{n+1,k+1}/I_{n,k}=1$ for any $n$ and $k$.}
\label{correction}
\end{center}
\end{figure}

Third, we show that the Laplace law of succession holds in QM up to an overall factor. 
More precisely, let us prepare an infinitely exchangeable state \cite{Caves02b} consisting of $n+1$ qubits and perform projective measurements $\{P, \openone-P\}$ to each constituent qubit, where $\openone$ is the identity operator of $\C^2$.
Then, given the set ${\cal M}_{n,k}$ of the outcomes that $P$ is true for $k$ times in the $n$ measurements, the conditional probability of $P$ being true for the $(n+1)$-th measurement takes the form of
\be
q(k_{n+1}|{\cal M}_{n,k})=\frac{k+1}{n+2}\frac{I_{n+1,k+1}}{I_{n,k}},
\label{q_ratio}
\ee
for three measures of quantum states we have investigated, {\it i.e.} the uniform measure on the pure state space, the flat measure, and the Bure measure \cite{Bengtsson06,Zyczkowski11}, all of which may represent our ignorance on the system.
Here, the overall factor $I_{n+1, k+1}/I_{n,k}$, hereafter called the correction term, is a polynomial of $n$ and $k$ determined by the measure we choose.
For example, the uniform measure on the pure state space yields $I_{n+1, k+1}/I_{n,k}=1$, suggesting that the eigenvalue distribution of the density operator contributes to the nontrivial correction term.
Note that the infinitely exchangeable state does not correspond with the maximal knowledge of the system, since it is not an i.i.d. array of a pure state in general.

Figure \ref{correction} shows that the correction terms tend to unity as the number of measurements $n$ increases with relative frequency $k/n$ fixed. Here we have used the flat measure and Bures measure.
Quantum conditional probability clearly approaches the Laplace law of succession in the large $n$ limit.
This implies that our degree of belief in terms of conditional probability obeys the correspondence principle, and can be evaluated with the relative frequency in a large number of the measurements.

The Laplace law of succession in QM implies that the rational degree of belief on a given quantum system can be approximated by the relative frequency.
As such, the rational degree of belief behaves as objective, since the relative frequency is uniquely determined, if the data are given.
This observation is in accordance with Earman\rq s assertion in \cite{pittphilsci14822} that the Lewis' Principal Principle, which states that subjective probabilities should be updated to match objective probabilities, and is derived as a theorem in QM.

As a technical remark, we point out that the correction term for the Bures measure yields a smaller deviation from the Laplace law of succession than that for the flat measure.
This is understood by the fact that the Bures measure is localized on the states of high purity \cite{Bengtsson06}, whence the behavior of the conditional probability should resemble that of the uniform measure on the pure state space, yielding the Laplace law of succession.

These results, as a whole, formally show that the probabilistic aspects of QM are subjective as much as probability theory, even taking the conditional probability and the L\"uders projection postulate into account. 
Nonetheless, we may employ the relative frequency even without invoking the maximal information of the system.
We hereafter give the technical details of the derivations, with a review of the arguments in classical probability theory.

\section{Arguments in Probability Theory}
\label{c_dba}
In the Bayesian viewpoint, probability is construed as a measure of our degree of belief on how a given statement is true under incomplete information.
To develop this idea precisely, the DBA was proposed by de Finetti \cite{deFinetti37,deFinetti92}.
Since then, the DBA has given rise to a myriad of succeeding research not only for classical probability \cite{Shimony55, Kemeny55, Lehman55, Gillies00} but also for QM \cite{Fuchs13, Caves02}.

In the DBA, our degree of belief is supposed to be measured using bets. 
We consider a bet played by a bettor and a bookmaker on whether a given proposition $a$ is true. 
 If the proposition is true, then the bettor obtains the stake $S$, which is a real number.
Otherwise, he obtains nothing.
To participate in the bet, the bettor pays the wager $qS$, where $q$ is the real number called the betting quotient.
The betting quotient $q$ plays a role as a measure of bettor\rq s degree of belief on how much the proposition is certain. 
Depending on the betting outcome, the stake is paid to the bettor. 
The payoff $G$ is given by $G=S-qS$ when the proposition is true, whereas $G=-qS$ when it is false.
The betting quotient, stake, and payoff with respect to the proposition $a$
are denoted by $q(a)$, $S(a)$ and $G(a)$, or $q_a$, $S_a$ and $G_a$ for short,
respectively.

Let us consider the case that the bettor and the bookmaker make bets on several propositions $a_1, a_2, \dots, a_n$. 
As before, the bettor assigns the betting quotient $q_i$ for the proposition $a_i$, pays the wager $q_iS_i$, and obtains the stake $S_i$ when the proposition $a_i$ is true. 
Then, the set of the betting quotients $q_1, q_2, \dots, q_n$ is called {\it coherent} \cite{deFinetti92, Shimony55} if the bettor wins at least one combination of the truth values of $a_i$. 
In other words, the payoff is non-negative with at least one combination of the truth values of $a_i$ for any sets of the stakes $S_i$.
In contrast, the set of the bets is called {\it Dutch book} if the bettor loses in any combinations of the betting outcomes.

We hereafter deal with the betting for the composite propositions.
Given the propositions, say $a_1$ and $a_2$, by using the Boolean operations
AND($\land$), OR($\lor$), NOT($\lnot$), we can make the composite
propositions $a_1\land a_2$, $a_1\lor a_2$, $\lnot a_i$, respectively.
Here the Boolean operations are defined by the following properties \cite{Whitesitt61}:
\begin{enumerate}
\item Commutativity
\be
a_1\land a_2=a_2\land a_1,
\qquad 
a_1\lor a_2=a_2\lor a_1.
\label{comm}
\ee
\item Distributivity
\be
a_1\land(a_2\lor a_3)&=&(a_1\land a_2)\lor(a_1\land a_3),\nonumber\\
a_1\lor(a_2\land a_3)&=&(a_1\lor a_2)\land(a_1\lor a_3).
\label{dist}
\ee
\item Identity
\begin{align}
a_1\land\Omega=a_1,
\qquad
a_1\lor\emptyset=a_1.
\label{boole_iden}
\end{align}
\item Complements
\begin{align}
a_1\land\lnot a_1=\emptyset,
\qquad
a_1\lor\lnot a_1=\Omega.
\label{boole_comp}
\end{align}
\end{enumerate}
Equations (\ref{boole_iden}) show that $\emptyset$ and $\Omega$ are the identities with respect to 
the OR operation and AND operation, respectively. 
We regard these identities as also propositions:
$\Omega$ is the always true proposition (tautology), whereas $\emptyset$ is the always false proposition (contradiction).
Note that $\lnot\Omega=\lnot\Omega\land\Omega=\emptyset,$ and 
$\lnot\emptyset=\lnot\emptyset\lor\emptyset=\emptyset\lor\lnot\emptyset=\Omega$, 
from Eqs.~(\ref{comm}), (\ref{boole_iden}), (\ref{boole_comp}).

Let the propositions $a_1, a_2,\dots, a_n$ be mutually exclusive and 
exhaustive: $a_i\land a_j=\emptyset$ for $i\neq j$, and 
$a_1\lor a_2\lor\dots\lor a_n=\Omega$. Let $A$ be the set whose elements
are $\emptyset, \Omega, a_1, a_2, \dots, a_n$, and all the composite
propositions thereof.
It has been shown in \cite{deFinetti37, deFinetti92, Shimony55, Kemeny55, Lehman55, Gillies00, Caves02} that if the set of  the betting quotients with respect to the elements of $A$ are coherent, then it satisfies the axioms of the probability: 
as a function of the propositions, the betting quotient $q(\cdot)$ is non-negative (\ref{positivity}), additive with respect to the mutually exclusive propositions (\ref{additivity}), and yields unity on the tautology $\Omega$ (\ref{tautology}). To make the description self-contained, we hereafter show these.

We first show the non-negativity (\ref{positivity}) of $q$, 
if the betting quotient $q$ is coherent.
This is shown as follows: Assume $q<0$. Then by setting $S<0$, the payoff is 
$G=S-qS<0$ for the proposition being true, and $G=-qS<0$ for that being false.
Thus, $q<0$ leads to the Dutch book situation. By taking the contraposition
of this observation, we can find Ineq.~(\ref{positivity}) if the betting quotient
is coherent.

Second, we prove the addition law of probability (\ref{additivity}),
for mutually exclusive (composite) propositions $a, b$.
To show this, we consider the bettings on the propositions $a, b$, and $c=a\lor b$. 
Depending on which combination of the propositions is found to be true,
the payoff is given as follows:
\be
G_a&=&(1-q_a)S_a - q_bS_b + (1-q_c)S_c    \quad\,\,\,\, \text{if $a$ is true}, \nonumber\\
G_b&=& - q_aS_a + (1-q_b)S_b + (1-q_c)S_c   \quad \text{if $b$ is true}, \nonumber\\
G_c&=&-q_aS_b - q_bS_b -q_cS_c      \qquad\qquad\quad \text{if $c$ is false}.
\label{payoff_additivity}
\ee
Two remarks are in order: First, $a\land b$ is always false, 
since the propositions $a$ and $b$ are mutually exclusive. Second, Eqs.~(\ref{payoff_additivity})
comprise a system of linear equations from the set of stakes to that of the possible payoffs. 

Now we come to the point. 
If Eqs.~(\ref{payoff_additivity}) have the solution as the linear system, the bookmaker is allowed to adjust his stakes in such a way that all the possible payoffs $G_a, G_b, G_c$ are negative, resulting in the Dutch book situation.
Taking the contraposition of this observation, we observe that the following must hold if the set of the betting quotients is coherent:
\be
\left|
\begin{array}{ccc}
1-q_a & -q_b & 1-q_c \\
-q_a & 1-q_b & 1-q_c \\
-q_a & -q_b & -q_c
\end{array}
\right|
=q_a+q_b-q_c
=0,
\ee
showing that Eq.~(\ref{additivity}) holds if the set of the betting quotient 
$\{q_a, q_b, q_c\}$ is coherent.

Third, we show that for the mutually exclusive and exhaustive propositions
$a_1, a_2, \dots, a_n$,
\be
q(a_1)+q(a_2)+\dots+q(a_n)=1
\label{totality}
\ee
holds if the betting quotients are coherent.
To prove this, we consider the bettings on the propositions $a_1, a_2, \dots, a_n$.
Since these propositions are mutually exclusive, the possible payoffs are given by
\be
G_i=(1-q_i)S_i - \sum_{j\neq i}q_jS_j   \quad \text{if $a_i$ is true},
\label{payoff_totality}
\ee
for all $i$. Here we have introduced abbreviated notation $q_i=q(a_i)$ and $S_i=S(a_i)$. 
Similarly to the derivation of Eq.~(\ref{additivity}), the bookmaker
can realize the Dutch book situation unless
\be
\left|
\begin{array}{cccc}
1-q_1 & -q_2 & \cdots & -q_n \\
-q_1 & 1-q_2 & \cdots & -q_n \\
\vdots & \vdots & \ddots & \vdots \\
-q_1 & -q_2 & \cdots & 1-q_n 
\end{array}
\right|
=1-\sum_{i=1}^nq_i
=0,
\ee
which leads to Eq.~(\ref{totality}). 
Furthermore, by using Eq.~(\ref{additivity}) and Eq.~(\ref{totality}),
we find
\be
q(\Omega)
&=&q(a_1\lor a_2 \lor \dots \lor a_n)\nonumber\\
&=&q(a_1) + q(a_2) + \dots + q(a_n)\nonumber\\
&=&1,
\label{identity}
\ee
which is one of the axioms of the probability (\ref{tautology}) announced before.
Moreover, since the betting quotients are non-negative, Eq.~(\ref{identity})
implies that $0\le q \le 1$ for all the betting quotients $q$, including those for the composite propositions. Together with Eqs.~(\ref{additivity}) and (\ref{identity}), 
we can conclude that the betting quotient obeys the axioms of probability.

So far, we have used the DBA to show that the coherent betting quotient is probability.
Not only this, we hereafter apply the DBA to argue the conditional probability.
For this, let us introduce conditional betting 
comprised of two propositions $a$ and $b$. In conditional betting,
the bookmaker first reveals whether $b$ is true. If $b$ is true,
the bookmaker proceeds to reveal whether $a$ is true.
If $b$ is false, the bet is called off and all the stakes and wagers associated with the conditioning by $b$ are returned.

On this regulation, suppose that we are given three bets: $b$, $a\land b$, and $a$ 
conditioned by whether $b$ is true. Let $q(a|b)$ and $S(a|b)$ be the betting 
quotient and stake associated with $a$ conditioned by $b$, respectively.
Further we introduce the shorthand notations $q=q(a\land b)$, $q^\prime=q(a | b)$, 
$q^{\prime\prime}=q(b)$ and $S=S(a\land b)$, $S^\prime=S(a | b)$, 
$S^{\prime\prime}=S(b)$. 
The possible payoffs are written as follows:
\be
G_1&=&(1-q)S + (1- q^\prime)S^\prime + (1-q^{\prime\prime})S^{\prime\prime} 
\,
\text{if $a\land b$ is true}, \nonumber\\
G_2&=& - qS - q^\prime S^\prime + (1-q^{\prime\prime})S^{\prime\prime}   \qquad\qquad\,\, 
\text{if $\lnot a \land b$ is true}, \nonumber\\
G_3&=&-qS - q^{\prime\prime}S^{\prime\prime}
\qquad\qquad\qquad\qquad\quad\,\,\,\,\,
\text{if $b$ is false}.
\label{payoff_conditioned}
\ee
As before, the bets become Dutch book unless
\be
\left|
\begin{array}{ccc}
1-q & 1-q^\prime & 1-q^{\prime\prime} \\
-q & -q^\prime & 1-q^{\prime\prime} \\
-q & 0 & -q^{\prime\prime} 
\end{array}
\right|
=q^\prime q^{\prime\prime} - q
=0,
\label{cond_coherence}
\ee
which results in the multiplication law of probability (\ref{c_cond}),
showing that the conditional betting quotient $q(a|b)$ is the conditional probability.
Note that
\be
q(a)=q(a|\Omega),
\label{prob_cond}
\ee
by setting $b=\Omega$ in Eq.~(\ref{c_cond}).

We now turn to the converse DBA \cite{Kemeny55, Lehman55, Gillies00}: 
In other words, we show that if the set of the betting quotients satisfies the axiom of the probability, then the betting quotients are coherent. 
This ensures that our degree of belief is identical to (conditional) probability if it is coherent.

To show this, let us give a general form of the payoff of the $n$ conditional bets.
Note that it suffices to consider the set of the conditional bets, since the outright bets are thought of as the conditional bets with tautology $\Omega$, due to Eq.~(\ref{prob_cond}).
Each conditional bet consists of the proposition $a_i$ conditioned by $b_i$, whose
betting quotient and stake are denoted by $q(a_i|b_i)$ and $S_i$, respectively.

Now consider $2^{2n}$ mutually exclusive composite propositions
\be
\omega&=&c_1\land\dots\land c_n\land d_1\land\dots\land d_n,
\ee
where
$c_i\in\{a_i, \lnot a_i\}$ and $d_i\in\{b_i, \lnot b_i\}$.
The proposition $\omega$ should be understood as a collective expression of the outcomes of the $n$ conditional bets.
Given the proposition $\omega$, then, the payoff of the conditional bets is written as
\be
G(\omega)=\sum_{i\in T(\omega)}\[1-q(a_i|b_i)\]S_i-\sum_{i\in F(\omega)}q(a_i|b_i)S_i,
\ee
where we have introduced the sets of indices $i$ such that
\be
T(\omega)&=&\{i\,|\, c_i=a_i \,\,{\rm and}\,\, d_i=b_i\},\nonumber\\
F(\omega)&=&\{i\,|\, c_i=\lnot a_i \,\,{\rm and}\,\, d_i=b_i\}.
\ee
While $a_i\land b_i$ is found to be true in the conditional bet $i\in T(\omega)$, $\lnot a_i\land b_i$ is found to be true in the conditional bet $i\in F(\omega)$. 

Following the standard procedure to show the converse DBA, we introduce the average payoff over the composite propositions $\omega$:
\be
\langle G \rangle =\sum_{\omega} q(\omega)G(\omega).
\label{ave_payoff}
\ee
Rewriting the average payoff as the function of the stakes, we obtain
\be
\langle G \rangle =x_1S_1+\dots+x_nS_n,
\label{Gexpanded}
\ee
where
\be
x_i=\[1-q(a_i|b_i)\]\sum_{\omega: i\in T(\omega)}q(\omega)-q(a_i|b_i)\sum_{\omega:i\in F(\omega)}q(\omega).\nonumber\\
\label{defxi}
\ee
Here the first sum is taken over $\omega$ on which the given proposition $a_i\land b_i$ is true, whereas the second sum is done over $\omega$ on which $\lnot a_i\land b_i$ is true.
The first sum reduces to $q(a_i\land b_i)$ since the elements of the set $\{\omega\}$ of the $2^{2n}$ propositions are mutually exclusive.
Similarly, the second sum reduces to $q(\lnot a_i\land b_i)$.
It follows from Eqs.~(\ref{dist}), (\ref{additivity}), and (\ref{c_cond}) that
\be
x_i&=&\[1-q(a_i|b_i)\]q(a_i\land b_i)-q(a_i|b_i)q(\lnot a_i\land b_i)\nonumber\\
&=&q(a_i\land b_i)-q(a_i|b_i)\[q(a_i\land b_i)+q(\lnot a_i\land b_i)\]\nonumber\\
&=&q(a_i\land b_i)-q(a_i|b_i)q(b_i)\nonumber\\
&=&0
\label{xi_zero}
\ee
for all $i$, which leads to $\langle G\rangle=0$. Since $q(\omega)\ge0$ and $\sum_\omega q(\omega)=1$, not all payoffs $G(\omega)$ are negative, showing coherence.
This completes the proof of the converse DBA.

Two remarks are in order. 
First, the proposition $a_i$ may be the same proposition as $a_j$ for some $j\neq i$, and $b_i$ is also the same as $b_j$.
This implies that the outcomes of the bets on a given proposition may change in the $n$ bets.
In other words, the outcome $c_i$ of the $i$-th bet may differ from $c_j$ even though $a_i=a_j$, and $d_i$ also may differ from $d_j$ even when $b_i=b_j$.
Second, we can complete the proof of the converse DBA even using 
\be
q^\prime(\omega)=\prod_{i=1}^nq(c_i|d_i)q(d_i)=\prod_{i=1}^nq(c_i\land d_i),
\ee
instead of $q(\omega)$.
The possible correlation among the outcomes of the mutually distinct conditional bettings in the joint probability $q(\omega)$ is irrelevant to show the converse DBA.
This alternative proof of the converse DBA shall be used in Sec.~\ref{q_dba}.

The arguments so far have shown that our rational degree of belief measured by the coherent betting quotient is equivalent to the probability.
Now we move to discuss how the rational degree of belief is related to the relative frequency, which provides another relevant interpretation of the probability: the frequency interpretation \cite{Gillies00}.
To argue this, let us consider a coin-flipping game for simplicity and introduce {\it infinite exchangeability} of the trials \cite{deFinetti92, Gillies00, Caves02b}, which is the prerequisite of the de Finetti representation theorem \cite{deFinetti92, Gillies00, Caves02b}.

Let $k_i=0,1$ be the outcome of the $i$-th coin flipping. 
The degree of belief $q(k_1,\dots, k_n)$ to obtain the outcomes $k_1,\dots, k_n$ is infinitely exchangeable if $q(k_1,\dots, k_n)$ is symmetric under all the permutations, that is,
\be
q(k_1,\dots, k_n)=q(k_{\sigma(1)},\dots, k_{\sigma(n)})
\ee
for any elements $\sigma\in\mathfrak{S}_n$ of the symmetric group $\mathfrak{S}_n$ and there exists a symmetric distribution $q(k_1,\dots,k_{n+m})$ such that 
\be
&&q(k_1,\dots, k_n)\nonumber\\
&&=\sum_{k_{n+1},\dots, k_{n+m}}q(k_1,\dots, k_n, k_{n+1},\dots,k_{n+m})
\ee
for all positive integer $m$. 
Then the de Finetti representation theorem states that there exists a probability measure $d\mu(p)$ such that
\be
q(k_1,\dots, k_n)=\int_{0}^1d\mu(p)p^k(1-p)^{n-k},
\ee
where $p$ is the probability for the outcome being $1$, and $k$ is the number of the events whose outcomes are $1$.
See \cite{Caves02b} for the proof.
The probability measure $d\mu(p)$ is a prior probability distribution representing our initial knowledge on the outcome distribution of the coin-flipping games.

It follows from the de Finetti representation theorem and Eq.~(\ref{c_cond}) that, given the set ${\cal M}_{n,k}$ of the outcomes that the $k$ heads are found in the $n$ tosses, the conditional probability distribution of $k_{n+1}=1$ takes the form
\be
q(k_{n+1}|{\cal M}_{n,k})=\frac{\int_{0}^1d\mu(p)p^{k+1}(1-p)^{n-k}}{\int_{0}^1d\mu(p)p^k(1-p)^{n-k}}.
\ee
Assuming that we have no prior knowledge of the outcome distribution of the coin-flipping game, we take the uniform distribution $d\mu(p)=dp$ as the prior.
We readily find
\be
q(k_1,\dots, k_n) = B(n-k+1,k+1),
\label{cq}
\ee
where $B(n, k)$ is the Beta function
\be
B(n,k)=\int_0^1dpp^{n-1}(1-p)^{k-1}.
\label{defBeta2}
\ee
Since $B(n,k)=(n-1)!(k-1)!/(n+k-1)!$, we then arrive at Eq.~(\ref{laplace}),
which is the Laplace law of succession \cite{Laplace02, Gillies00, deFinetti37, deFinetti92, Cox46, Cox61}. 
Thus, in large $n$ and $k$, the degree of belief $q(k_{n+1}|{\cal M}_{n,k})$ approaches the relative frequency $k/n$ under no prior knowledge situation.


\section{Quantum Dutch Book Argument and its Converse}
\label{q_dba}

In this section, we extend the argument given in Sec.~\ref{c_dba} to QM.
We shall find that the quantum probability, as well as conditional probability, are derived from the DBA, and the converse DBA holds.
These suggest that our coherent degree of belief on the quantum system is equivalent to quantum (conditional) probability.

Let ${\cal H}$ be a $d$-dimensional Hilbert space ($d<\infty$), and $P$ be a projector acting on ${\cal H}$. 
${\cal L}({\cal H})$ denotes the set of all the projectors acting on ${\cal H}$. 
Every projector can be treated as a proposition because the eigenvalues of the projectors are either 1 or 0, seen as the truth values.
In the same way as Sec.~\ref{c_dba}, we introduce the betting quotient $q(P)$, which measures our degree of belief for the proposition $P$ being true on a quantum system in ${\cal H}$ under consideration.

Suppose that we are given the propositions $P$ and $Q$. 
Then the composite projectors $P\land Q$ and $P\lor Q$ are defined as the projectors to the meet and join of the range of $P$ and that of $Q$, respectively. 
The negation $\lnot P$ is the projector on the orthogonal complement of the range of $P$. In particular, if $P$ and $Q$ commute, then
\be
P\land Q=PQ,
\quad
P\lor Q= P+Q-PQ,
\quad
\lnot P=\mathbbm{1}-P,\nonumber\\
\label{qBoole_op}
\ee
where $\mathbbm{1}$ is the identity operator on ${\cal H}$. 
For the commuting projectors $P, Q$ such that $[P, Q]=0$, we define $P\le Q$, which means $PQ=QP=P$.

To proceed, we hereafter assume that the betting quotient and coherence are well-defined for the bettings comprised of commuting projectors. The DBA in Sec.~\ref{c_dba} is then readily applied to the bets made of the commuting projectors.
Indeed, the Born rule was derived in \cite{Caves02} as follows: We derive 
\be
&&q(P)\ge0,
\qquad
q(\openone)=1,\nonumber\\ 
&&q(P\lor Q)=q(P)+q(Q)
\label{q_prob}
\ee 
 for the mutually exclusive commuting projectors such that $P\land Q=PQ=0$, by repeating the DBA with formal replacement of the propositions in the previous section by the projectors. Since Eqs.~(\ref{q_prob}) are sufficient conditions of the Gleason theorem \cite{Gleason57}, we obtain the Born rule
\be
q(P)=\tr(\rho P),
\ee
where $\rho$ is a density operator acting on ${\cal H}$.

In parallel with the Born rule, we can derive a concrete expression of the degree of belief for conditional betting on the projectors.
We begin with defining the conditional betting of $P$ conditioned by $Q$ in a way similar to the classical one, even though $P$ and $Q$ do not commute: The bookmaker first measures $Q$ on a quantum system and reveals whether $Q$ is true.
If $Q$ is true, the bookmaker further measures $P$ on the system after the measurement of $Q$ and reveals whether $P$ is true. If $Q$ is false, the bet is called off as in the case of classical conditional betting.
We denote by $S(P|Q)$ the stake of the bet on $P$ conditioned by $Q$.

Let $q(P|Q)$ be the degree of belief for the projector $P$, given that $Q$ is true.
We now make two assumptions: First, $q(P|Q)$ is well-defined even for not necessarily commuting projectors $[P, Q]\neq0$.
Second, similarly to $q(\cdot)$ for the outright bet, $q(\cdot|Q)$ is coherent for the conditional bettings comprised of commuting projectors conditioned by another projector $Q$.
Note that the projector $Q$ used for the conditioning is not necessarily commuting with $P$.
For example, given two conditional bettings, $P_1$ conditioned by $Q$ and $P_2$ conditioned by $Q$, we require coherence only when $P_1$ and $P_2$ commute.

On the basis of these assumptions, it is straightforward to show that the degree of belief $q(P|Q)$ is the probability distribution when $Q$ is fixed.
More precisely, $q(P|Q)$ satisfies i)
$
q(P|Q)\ge0,
$
ii)
$
q(\openone|Q)=1
$
for any pairs of $P, Q$,
and iii) the additivity
\be
q(P\lor Q |R)=q(P|R)+q(Q|R),
\label{q_cond_prob}
\ee
 for any projectors $P, Q, R$ such that $P\land Q=PQ=0$.

Similarly to the derivation of Ineq.~(\ref{positivity}), let us show $q(P|Q)\ge0$: 
Consider the conditional bet on $P$ conditioned by $Q$ and suppose that the proposition $Q$ is found to be true. 
If the proposition $P$ is also found to be true and $q(P|Q)<0$, the payoff is given as $S(P|Q)-q(P|Q)S(P|Q)$, which is negative when the bookmaker chooses $S(P|Q)<0$.
On the other hand, if the proposition $P$ is also found to be false and $q(P|Q)<0$, the payoff is given as $-q(P|Q)S(P|Q)$, which is also negative when the bookmaker chooses $S(P|Q)<0$.
Thus, the payoff is always negative regardless of whether $P$ is true or false, suggesting the Dutch book.

Next, we show the additivity $q(P\lor Q|R)=q(P|R)+q(Q|R)$ for the projectors such that $PQ=0$: Consider the conditional bets $P$, $Q$ and $P\lor Q$ given the condition $R$. Since all the conditional bets are called off if $R$ is false, it suffices to consider the case that  $R$ is found to be true, where the payoffs take the form of Eqs.~(\ref{payoff_additivity}) with suitable replacements of the indices, and thereby the coherence condition yields the law of the additivity for the conditional bets. 
Note that $q(\openone|Q)=1$ is also derived in the same manner.
These properties of the conditional betting quotient show that $q(P|Q)$ is the probability distribution when $Q$ is fixed, as in the case of the outright betting quotient $q(P)$.

Having found that $q(\cdot|Q)$ is the probability distribution, we consider three bets consisting of the commuting projectors $P$ and $Q$: $Q$, $P\land Q=PQ$, and $P$ conditioned by whether $Q$ is true.
Since $P$ and $Q$ commute, the argument in the previous section can be applied to the present bets, and thereby we obtain
\be
q(P\land Q)=q(P|Q)q(Q)
\label{q_com_multi}
\ee
for any pairs of commuting projectors.
Note that
\be
q(P|Q)=\frac{q(P)}{q(Q)},
\quad
{\rm if}\,\,
P\le Q\,\,
{\rm and}\,\,
q(Q)\neq0,
\label{ordering}
\ee
since $P\le Q$ implies $P\land Q=PQ=P$.

With Eq.~(\ref{ordering}) for the degree of belief $q(P|Q)$, we employ the following lemma:
\begin{lemma}[Beltrametti and Cassinelli \cite{BC81, Malley04, Ichikawa18}]
\label{BC_lemma}
For Hilbert space ${\cal H}$, let $q(\cdot)$ be a probability measure on ${\cal L}({\cal H})$, and let $Q$ be any projector such that $q(Q)\neq0$. Then there exists a unique probability measure on ${\cal L}({\cal H})$, which we denote by
$q(\cdot| Q)$, such that for all projectors $P\le Q$ it is the case that $q(P|Q) =q(P)/q(Q)$.
\end{lemma}
This lemma ensures the uniqueness of the degree of belief $q(P|Q)$ as the conditional probability distribution.
Due to the additivity $q(P\lor Q|R)=q(P|R)+q(Q|R)$ for the projectors such that $PQ=0$, we apply the Gleason theorem to $q(\cdot | Q)$, and find that there exists an appropriate density operator $\rho_Q$ with which the coherent degree of belief takes the form of Eq.~(\ref{q_cond}).
Since $q(Q|Q)=1$ from Eq.~(\ref{ordering}), we find that $\tr(\rho_Q Q)=1$, which suggests that the density operator $\rho_Q$ has non-vanishing support only on the range of $Q$.
Hence, there exists a density operator $\rho$ such that $\rho_Q$ takes the form (\ref{rhoq}),
from which we obtain the concrete expression of the coherent degree of belief
\be
q(P|Q)=\frac{\tr(Q \rho QP)}{\tr(\rho Q)}.
\ee
As stated in Sec.~\ref{OR}, our derivation of quantum conditional probability leads to the Bayesian interpretation of the L\"uders projection postulate.

The argument developed so far is the quantum analogue of the DBA, showing that our degree of belief on what occurs in quantum systems takes the forms of the Born rule as well as the quantum conditional probability (\ref{q_cond}) on the assumption of the coherence of the (conditional) betting quotients of commuting projectors. 
We can further show the converse DBA in QM: the Born rule and the quantum conditional probability (\ref{q_cond}) lead to the coherence of the betting quotients.

The proof is completely parallel to the argument used for classical probability: 
As in Sec.~\ref{c_dba}, we begin with a general form of the payoff of the $n$ conditional bets on projectors.
Since $q(P|\openone)=q(P)$, it is sufficient to consider the set of conditional bets.
Each conditional bet consists of the proposition $P_i$ conditioned by $Q_i$, whose
betting quotient and stake are denoted by $q(P_i|Q_i)$ and $S_i$, respectively.

Let $w$ be a combination of the outcomes of the $n$ conditional bets such that
\be
\!\! w=(v_1(P_1), \dots, v_n(P_n), v_{n+1}(Q_1), \dots, v_{2n}(Q_n)),
\ee 
where we have introduced the outcome
\be
v_i(P)\in\{0,1\},
\ee
for $i=1,\dots, 2n$.
Associated with this, we define
\be
\tau_i(P)=\begin{cases}
 P     & \text{if\,\,\,} v_i(P)=1, \\
 \lnot P     & \text{if\,\,\,} v_i(P)=0.
\end{cases}
\ee

Given the combination $w$, then, the payoff of the conditional bets is written as
\be
G(w)=\sum_{i\in T(w)}\[1-q(P_i|Q_i)\]S_i-\sum_{i\in F(w)}q(P_i|Q_i)S_i, \nonumber\\
\ee
where we have introduced the sets of indices $i$ such that
\be
T(w)&=&\{i\,|\, v_i(P_i)=1 \,\,{\rm and}\,\, v_{n+i}(Q_i)=1\},\nonumber\\
F(w)&=&\{i\,|\, v_i(P_i)=0 \,\,{\rm and}\,\, v_{n+i}(Q_i)=1\},
\ee
which are analogues of $T(\omega)$ and $F(\omega)$ in Sec.~\ref{c_dba}.

We take the average of the payoff $G(w)$ over the combinations of the outcomes of the conditional bets $w$:
\be
\langle G \rangle_{\rm QM} =\sum_{w} q(w)G(w),
\label{q_ave_payoff}
\ee
where
\be
q(w)=\prod_{i=1}^n q(\tau_i(P_i)|\tau_{n+i}(Q_i))\, q(\tau_{n+i}(Q_i))
\ee
is the probability to obtain the outcomes $w$ in the $n$ conditional bets.
Indeed, we observe
\be
q(w)\ge0,
\qquad
\sum_wq(w)=1
\label{qw}
\ee
from
\be
q(P|Q)+q(\lnot P|Q)=1,
\qquad
q(P)+q(\lnot P)=1.
\label{q_identity}
\ee

We rewrite the average payoff as the function of the stakes, and obtain
\be
\langle G \rangle_{\rm QM} =x_1S_1+\dots+x_nS_n,
\label{q_Gexpanded}
\ee
where
\be
x_i=\[1-q(P_i|Q_i)\]\sum_{w: i\in T(w)}q(w)
-q(P_i|Q_i)\sum_{w:i\in F(w)}q(w).\nonumber\\
\label{q_xi}
\ee
Here the first sum is over $w$ with $v_i(P_i)=v_{n+i}(Q_i)=1$, whereas the second sum is over $w$ with $v_i(P_i)=0$ and $v_{n+i}(Q_i)=1$.
Since
\be
\sum_{w: i\in T(w)}q(w)&=&q(P_i|Q_i)q(Q_i),\nonumber\\
\sum_{w: i\in F(w)}q(w)&=&q(\lnot P_i|Q_i)q(Q_i),
\label{q_marginal}
\ee
from Eqs.~(\ref{q_identity}),
by plugging Eqs.~(\ref{q_marginal}) into Eq.~(\ref{q_xi}), we observe
\be
x_i&=&\[1-q(P_i|Q_i)\]q(P_i|Q_i)q(Q_i)\nonumber\\
&&-q(P_i|Q_i)q(\lnot P_i|Q_i)q(Q_i)\nonumber\\
&=&\[1-q(P_i|Q_i)-q(\lnot P_i|Q_i)\]q(P_i|Q_i)q(Q_i)\nonumber\\
&=&0
\label{q_xi_zero}
\ee
for all $i$, from Eqs.~(\ref{q_identity}).
This leads to $\langle G\rangle_{\rm QM}=0$. Equations (\ref{qw}) then imply that not all payoffs $G(w)$ are negative, showing coherence.
This shows that the converse DBA holds in QM.

\section{Connection to Relative Frequency}
\label{q_rel_freq}

We turn to discuss the relation between the conditional degree of belief $q(P|Q)$ and relative frequency.
Similarly to classical probability, we argue this with the measurements of two-level systems, by introducing the quantum de Finetti representation theorem \cite{Caves02b}, which requires exchange symmetry on the density operators.

Consider a tensor product Hilbert space ${\cal K}^n={\cal H}^{\otimes n}$ with ${\cal H}=\mathbbm{C}^N$.
Given a permutation $\sigma$ of the symmetric group $\mathfrak{S}_n$, let us suppose that there exists a self-adjoint operator $\pi_\sigma$ such that
\be
\pi_\sigma\Bot_{i=1}^n\ket{e_i}_i=\Bot_{i=1}^n\ket{e_i}_{\sigma(i)},
\ee
where $\{\ket{e_i}_i\}$ is a complete orthonormal basis of the $i$-th constituent Hilbert space. We define the totally symmetric subspace of the tensor product Hilbert space ${\cal K}^n$ as
\be
{\cal K}_{\cal S}^n=\{\ket{\psi}\in{\cal K}^n\,|\, \pi_\sigma\ket{\psi}=\ket{\psi}, \, \forall\sigma\in\mathfrak{S}_n\}.
\ee

Let $\rho_n$ be a density operator whose non-vanishing support is only on ${\cal K}_{\cal S}^n$.
We call such a density operator a symmetric density operator.
The symmetric density operator $\rho_n$ is {\it infinitely exchangeable} if there exists a symmetric density operator $\rho_{n+m}$ acting on ${\cal K}^{n+m}$ such that
\be
\rho_n=\tr_m(\rho_{n+m}),
\ee
for all positive integer $m$.
Here, $\tr_m(\cdot)$ denotes the partial trace operation with respect to an $m$-partite subsystem.
Note that the choice of the subsystem to be traced out is irrelevant, due to the exchange symmetry of $\rho_{n+m}$.

Quantum de Finetti representation theorem states that given an infinitely exchangeable density operator $\rho_n$, there exists a density operator $\rho$ acting on ${\cal H}$ such that \be
\rho_n=\int d\mu(\rho)\rho^{\otimes n},
\label{q_deFinetti_rep}
\ee
where $\mu(\rho)$ is a probability measure of the density operator $\rho$. 
Now we introduce the projector
\be
P_{n,k}=P^{\otimes k}\otimes (\lnot P)^{\otimes (n-k)}
\ee
and define
the probability distribution of observing the projector $P$ being true for the first $k$ times in $n$ repeated measurements by
\be
q(P_{n,k})
=\tr(\rho_n P_{n,k})
=\int d\mu(\rho)q(P)^k(1-q(P))^{n-k}.\nonumber\\
\label{qpro}
\ee
Furthermore, by following the notation in Sec.~\ref{c_dba}, we denote  by $q(k_{n+1}|{\cal M}_{n,k})$ the conditional probability of the $(n+1)$-th measurement of $P$ being true, given the set ${\cal M}_{n,k}$ of the outcomes that $P$ is true for $k$ times in the $n$ measurements. 
Since the ordering of the outcomes occurring is immaterial in this case, the projection operator associated with ${\cal M}_{n,k}$ is given by
\be
\Pi_{n,k}&=&P_{n,k}+(\lnot P)\otimes P_{n-1,k}+P_{2,1}\otimes P_{n-2, k-1}+\cdots\nonumber\\
&&+(\lnot P)^{\otimes(n-k)}\otimes P^{\otimes k},
\label{defpi}
\ee
consisting of the $\binom{n}{k}$ mutually orthogonal projectors.
With the projector $\Pi_{n,k}$, the conditional probability $q(k_{n+1}|{\cal M}_{n,k})$ takes the form of
\be
q(k_{n+1}|{\cal M}_{n,k})
&=&q(P\otimes\openone^{\otimes n}|\openone\otimes \Pi_{n,k})\nonumber\\
&=&\frac{q(P_{n+1,k+1})}{q(P_{n,k})},
\label{qcond}
\ee
where we have employed the density operator $\rho_{n+1}$.
 For the derivation of Eq.~(\ref{qcond}), see Appendix \ref{der_qcond}.

The classical uniform probability measure leads to the Laplace law of succession (\ref{laplace}).
For comparison, we wish to employ concrete expressions of the uniform measures on the density operators.
Notably enough, it has been known that the metric of the space of the density operators is not uniquely determined in general  \cite{Bengtsson06}.
For example, there exists a family of infinite numbers of monotone metrics with respect to the completely positive and trace-preserving (CPTP) operations.
Taking into account this non-uniqueness feature, we shall consider the three measures as possible substitutes for the classical uniform measure: the uniform measure on the pure state space, flat measure, and Bures measure.
Note that the flat measure and Bures measure are those of the space of the density operators.

To deal with the three measures in a unified manner, let us follow \cite{Zyczkowski11} and introduce a family of the measures called product measures by paying attention to the spectral decomposition of the density operator
\be
\rho=U\Lambda U^\dagger,
\qquad
\Lambda={\rm diag}(\lambda_1,\dots,\lambda_N),
\label{decrho}
\ee
where $U$ is a unitary operator of $SU(N)$ and $\lambda_i$ are the eigenvalues of $\rho$ such that
\be
\lambda_1+\dots+\lambda_N=1,
\qquad
\lambda_i\ge0,
\label{defLambda}
\ee
for all $i=1,\dots,N$.
This decomposition enables us to see the set of the density operators as a direct product of $SU(N)$ and the set of the eigenvalues $\{\lambda_i\}$ satisfying Eq.~(\ref{defLambda}).
Accordingly, we may construct the measures of the density operators as the product of the measure $dV$ of $SU(N)$ elements and that of $\{\lambda_i\}$ denoted by $dF$, {\it i.e.},
\be
d\mu(\rho)=dVdF,
\ee
which is the product measure announced before.
As we shall see, all the measures aforementioned are of the family of the product measures.

We hereafter work with a system comprised of the spins ${\cal H}=\mathbbm{C}^2$ and consider the measurements of the projector
\be
P=\ket{\uparrow}\bra{\uparrow},
\ee
where 
$
\ket{\uparrow}=(
\begin{smallmatrix}
1\\
0
\end{smallmatrix}
)
$.
This choice of the measurement is sufficient, since the measures we hereafter consider are constructed with the Haar measure, and thereby invariant under the action of any elements of the unitary group.
We employ the spectral decomposition (\ref{decrho}) and use the Euler angle representation of $SU(2)$, that is, for $U \in SU(2)$,
\be
U=e^{i\alpha\sigma_3}e^{i\beta\sigma_2}e^{i\gamma\sigma_3},
\label{euler}
\ee
with
$
\sigma_2=
(\begin{smallmatrix}
 0     & -i   \\
 i     &  0
\end{smallmatrix})
$
and
$
\sigma_3=
(\begin{smallmatrix}
 1     & 0   \\
 0     &  -1
\end{smallmatrix})
$.
Here, the ranges of the parameters $\alpha, \beta, \gamma$ are $0\le\alpha, \gamma\le\pi$, and $0\le\beta\le\pi/2$, respetively.
In the Euler angle representation, we immediately find
\be
q(P)=\lambda_1\cos^2\beta+\lambda_2\sin^2\beta.
\label{qp}
\ee
In addition to this, we employ the Haar measure of $SU(2)$ group as $dV$, which takes the form of
\be
dV= \frac{\sin2\beta}{\pi^2}d\alpha d\beta d\gamma,
\label{Haar}
\ee
and we may write
\be
dF=F(\lambda_1, \lambda_2)d\lambda_1d\lambda_2,
\ee
in view of the decomposition (\ref{decrho}) \cite{Tilma02a, Tilma02b}.
The choice of the measure $F(\lambda_1, \lambda_2)$ leads to three distinct measures mentioned above.

\subsection{Uniform measure on pure state space}
We begin with the product measure with
\be
F(\lambda_1, \lambda_2)=\delta(\lambda_1-1)\delta(\lambda_1+\lambda_2-1),
\label{dirac}
\ee
where $\delta(x)$ is the Dirac delta function.
The measure $F(\lambda_1, \lambda_2)$ has non-vanishing support only on pure states.
Since the Haar measure $dV$ is invariant under the action of $SU(2)$, $d\mu(\rho)$ with Eq.~(\ref{dirac}) is the uniform measure on the space of the pure states.
By performing the integration in Eq.~(\ref{qpro}) as shown in Appendix \ref{app_pure}, we readily find 
\be
q(P_{n,k})=B(n-k+1, k+1),
\label{q_pure}
\ee
which leads to the Laplace law of succession (\ref{laplace}).

\subsection{Flat measure}
We next turn to the flat measure, which is the product measure composed of
the Haar measure (\ref{Haar}) and
\be
F(\lambda_1, \lambda_2)=\delta(\lambda_1+\lambda_2-1),
\label{Fflat}
\ee
indicating uniformity with respect to the eigenvalue distribution.
The probability associated with the flat measure takes the form of
\be
q(P_{n,k})=B(n-k+1, k+1)I_{n,k},
\label{mixed}
\ee
with
\be
I_{n,k}&=&\sum_{j=0}^k\sum_{l=0}^{n-k}\binom{j+l}{j}\binom{n-j-l}{k-j}
B(n-r+1, r+1),\nonumber\\
\label{ink_flat}
\ee
where we have introduced
\be
r=k-j+l.
\ee
For the derivation of Eq.~(\ref{ink_flat}), see Appendix \ref{app}.
The conditional probability is, accordingly, given by Eq.~(\ref{q_ratio}).

\subsection{Bures measure}
The Bures measure \cite{Bengtsson06, Zyczkowski11} is also the product measure consisting of the Haar measure and
\be
F(\lambda_1,\lambda_2)=\frac{2}{\pi}\frac{(\lambda_1-\lambda_2)^2}{\sqrt{\lambda_1\lambda_2}}\delta(\lambda_1+\lambda_2-1).
\label{FBures}
\ee
The Bures measure is derived from the Bures metric, which is a natural metric equipped in the space of the density operators \cite{Bengtsson06}.
The distance with respect to the Bures metric is called the Bures distance, which is the function of fidelity.
Since fidelity is a measure of the distinguishability of the density operators, the Bures measure has a clear operational meaning as the measure of state discrimination.

The probability distribution $q(P_{n,k})$ and conditional probability pertaining to the Bures measure take the forms of Eq.~(\ref{mixed}) and Eq.~(\ref{q_ratio}), respectively, with
\be
I_{n,k}&=&\frac{2}{\pi}\sum_{j=0}^k\sum_{l=0}^{n-k}\binom{j+l}{j}\binom{n-j-l}{k-j}\nonumber\\
&&\times\frac{(n-2r)^2+n+1}{(r+\frac{1}{2})(n-r+\frac{1}{2})}B(n-r+\frac{3}{2}, r+\frac{3}{2}),
\label{ink}
\ee
which is derived in Appendix \ref{app_Bures}.

\section{Conclusions and Discussion}
\label{conc}

We have examined whether the quantum probability, as well as conditional probability, can be construed as our degree of belief on the quantum systems satisfying the coherence condition.
Our finding is the following: exactly the same as classical probability, the quantum probability is identical to the degree of belief with coherence.
More precisely, our degree of belief with coherence can take the form of the Born rule as well as the quantum conditional probability (\ref{qcond}), and conversely, they fulfill the coherence condition to any combinations of the (conditional) bettings.
Hence, we can understand any quantum mechanical descriptions of the quantum systems in terms of Bayesian notions.

The coincidence between quantum probability and the coherent degree of belief should be distinguished from other major interpretations of classical probability.
For example, in the logical interpretation of classical probability, the (conditional) probability is equivalent to the measure of the plausibility of a logical statement under incomplete information \cite{Kaynes21, Cox1946_nature,Cox46, Cox61, Jaynes03, vanHorn03}, which obeys natural inference rules. In contrast, quantum conditional probability does not obey the inference rules in the case of non-commuting projectors \cite{Ichikawa18}.
Bearing the above case in mind, we may say that this work is the first derivation of quantum conditional probability and the post-measurement state (\ref{rhoq}) in a way consistent with the epistemic viewpoint of the probability theory.

Our derivation of the post-measurement state (\ref{rhoq}) associated with the quantum conditional probability is complementary to the derivation of the (unnormalized) post-measurement state (\ref{ave_rhoq}) in \cite{Fuchs_Schack2012} for the case of the outcome aggregation.
Whereas we have derived the state (\ref{rhoq}) by using the quantum analogue of the DBA,  \cite{Fuchs_Schack2012} has derived the state (\ref{ave_rhoq}) by considering the {\it diachronic} DBA \cite{Vineberg2011} on the assumption of the L\"uders rule $\rho\rightarrow\rho_Q$ without the outcome aggregation.
Here the diachronic DBA is a variant of the DBA in such a way as to take into account the time development of the degree of belief.
Therefore our derivation may support the assumption employed in \cite{Fuchs_Schack2012}.

In addition to the investigation of the above foundational issues, we ensured that quantum probability approaches the relative frequency in large $n$ limit, through the Laplace law of succession with the correction terms.
Our argument is perfectly in parallel with that developed by de Finetti for probability theory, and thereby directly comparable with it.
Note that the i.i.d. pure states are the special cases of infinitely exchangeable states.
Thus, our result supports the result in \cite{Caves02} on the ground of the weaker assumption.

Here we point out the resemblance between the Laplace law of succession (in QM) and the concept of the convergence to the truth, proposed by the pragmatist Peirce, which is summarized in the following quote:
\begin{quote}
Different minds may set out with the most antagonistic views, but the progress of investigation carries them by a force outside of themselves to one and the same conclusion \cite{Peirce1878}.
\end{quote}
In parallel with this convergence, the Laplace law of succession ensures that the degrees of belief of all the rational agents converge approximately to the relative frequency, through the repetition of the observations.
This resemblance may show the affinity between QBism and the pragmatism not only of James and Dewey \cite{Fuchs2017} but also of Peirce.

It is interesting to mention the relation between our work presented here and an ontological viewpoint on QM: Pusey, Barrett, Rudolph (PBR) theorem \cite{Pusey12, Leifer14} is a no-go theorem that any epistemic descriptions properly formulated in terms of some hidden variables contradict with QM, on the assumption of independent preparation of the system.
Moreover, it is pointed out in \cite{Lewis12, Leifer14} that without the independent preparation assumption, we construct epistemic models consistent with QM.
Bearing these works in mind, it is notable that our analysis is based on two assumptions: our degree of belief is measurable using bets in principle, and the propositions on the quantum systems are described by the projection operators.
Since the Bayesian probability is an epistemic notion and the assumptions we have used seem weaker than those of the PBR theorem, our result may support the counterexample to the PBR theorem without invoking the hidden variable description.

We point out that our study can be applicable to quantum-like modeling of the process of decision making and cognitive studies \cite{doi:10.1098/rsta.2015.0245,e25060886}, where update of one\rq s belief is of central concern. 
In our work, the post-measurement state $\rho_Q$ and its aggregated form $\rho_{\langle Q\rangle}$ naturally appear.
Therefore our work ensures the use of the state update rule associated with $\rho_Q$ and $\rho_{\langle Q\rangle}$ in quantum-like modeling.

In closing, we turn to the subject-object distinction in QM, which is one of the roots of our investigation.
On the basis of our results, the objective aspects of QM lie not on the Born rule and the L\"uders projection postulate, but on the other axioms unexplored here.
In particular, to use Jaynes\rq~phrase, it could be of interest to ask whether we can \lq\lq separate the subjective and objective aspects\rq\rq~in our assumption that the projectors represent the propositions on the quantum system, since the contextuality \cite{Koc,RevModPhys.94.045007}, one of the genuine quantum phenomena, is the direct consequence of the non-Boolean lattice structure of the projectors. 
We close this paper by quoting Heisenberg\rq s phrase again, which could imply the subject-object coexistence in the algebraic structure of the projectors:

\begin{quotation}
[..., and] we have to remember that what we observe is not nature in itself but nature exposed to our method of questioning. Our scientific work in physics consists in asking questions about nature in the language that we possess and trying to get an answer from experiment by the means that are at our disposal. \cite{Heisenberg1958-HEIPAP}
\end{quotation}

\section*{Acknowledgment}
We thank S. Tanimura, G. Kimura, and H. Hakoshima for fruitful discussions.
This work was supported by MEXT Quantum Leap Flagship Program (MEXT Q-LEAP) Grant Number JPMXS0120319794.

\appendix
\section{Proof of Eq.~(\ref{qcond})}
\label{der_qcond}
To derive Eq.~(\ref{qcond}), we first note that
\be
\Pi_{n,k}^2=\Pi_{n,k},
\ee
from which we find
\be
q(P\otimes\openone^{\otimes n}|\openone\otimes \Pi_{n,k})
=\frac{\tr(\rho_{n+1}P\otimes\Pi_{n, k})}{\tr(\rho_{n+1}\openone\otimes \Pi_{n,k})}.
\ee
For the numerator, we find
\be
\!\!\tr(\rho_{n+1}P\otimes\Pi_{n, k})
&=&\int d\mu(\rho)\tr[(\rho P)\otimes(\rho^{\otimes n}\Pi_{n,k})]\nonumber\\
&=&\int d\mu(\rho)\tr(\rho P)\tr(\rho^{\otimes n}\Pi_{n,k}).
\label{ppi}
\ee
It follows from the exchange symmetry of the density operator $\rho^{\otimes n}$ that
\be
\tr(\rho^{\otimes n}\Pi_{n,k})=\binom{n}{k}\tr(\rho^{\otimes n}P_{n,k}).
\label{pip}
\ee
Plugging this expression to Eq.~(\ref{ppi}), we obtain
\be
\tr(\rho_{n+1}P\otimes\Pi_{n, k})
&=&\binom{n}{k}\int d\mu(\rho)\tr(\rho P)\tr(\rho^{\otimes n}P_{n,k})\nonumber\\
&=&\binom{n}{k}\int d\mu(\rho)\tr[(\rho P)\otimes(\rho^{\otimes n}P_{n,k})]\nonumber\\
&=&\binom{n}{k}\tr(\rho_{n+1}P_{n+1,k+1})\nonumber\\
&=&\binom{n}{k}q(P_{n+1,k+1}).
\ee

Similarly to the numerator, we enumerate the denominator as follows.
\be
\tr(\rho_{n+1}\openone\otimes\Pi_{n, k})
&=&\int d\mu(\rho)\tr[\rho\otimes(\rho^{\otimes n}\Pi_{n,k})]\nonumber\\
&=&\int d\mu(\rho)\tr(\rho)\tr(\rho^{\otimes n}\Pi_{n,k})\nonumber\\
&=&\int d\mu(\rho)\tr(\rho^{\otimes n}\Pi_{n,k}).
\label{denominator}
\ee
By using Eq.~(\ref{pip}), we observe
\be
\tr(\rho_{n+1}\openone\otimes\Pi_{n, k})
&=&\binom{n}{k}\int d\mu(\rho)\tr(\rho^{\otimes n}P_{n,k})\nonumber\\
&=&\binom{n}{k}\tr(\rho_nP_{n,k})\nonumber\\
&=&\binom{n}{k}q(P_{n,k}),
\label{denom1}
\ee
which completes the proof. 

Note that Eq.~(\ref{pip}) reads an instance of the principle of indifference \cite{Laplace02, Jaynes03}; when the ordering of the outcomes is irrelevant, it is reasonable to assign the probability for a possible ordering of the outcomes by evenly distributing the value of the original probability.

\section{Derivation of Eq.~(\ref{q_pure})}
\label{app_pure}
To compute $q(P_{n,k})$ with the measure (\ref{dirac}), we first perform the integration with respect to $\lambda_1, \lambda_2$, which results in
\be
q(P_{n,k})
&&=\iiint_\mathcal{R} d\alpha d\beta d\gamma 
\frac{\sin2\beta}{\pi^2}(\cos\beta)^{2k}(\sin\beta)^{2(n-k)},
\nonumber\\
\ee
where $\mathcal{R}=\{(\alpha, \beta, \gamma)\,|\, 0\le\alpha, \gamma\le\pi,\,\, 0\le\beta\le\pi/2\}$.
We further integrate $\alpha$ and $\gamma$ to obtain
\be
q(P_{n,k})=
2\int_0^{\frac{\pi}{2}} d\beta (\sin\beta)^{2(n-k)+1}(\cos\beta)^{2k+1}.
\ee
By comparing this expression to the definition of the Beta function
\be
B(n,k)=
2\int_0^{\frac{\pi}{2}} d\beta (\sin\beta)^{2n-1}(\cos\beta)^{2k-1},
\label{defBeta1}
\ee
we obtain Eq.~(\ref{q_pure}).

\section{Derivation of Eqs.~(\ref{ink_flat})}
\label{app}
We first perform the integration with respect to $\alpha$ and $\gamma$, which leads to
\be
q(P_{n,k})=\int_0^1d\lambda_1\int_0^1d\lambda_2F(\lambda_1,\lambda_2)J_{n,k},
\label{target}
\ee
where
\be
J_{n,k}=\int_0^{\frac{\pi}{2}}d\beta\sin2\beta q(P)^k(1-q(P))^{n-k}.
\label{jnk}
\ee

Let us next evaluate $J_{n,k}$. 
The binomial theorem and Eq.~(\ref{qp}) lead to
\be
q(P)^k
=\sum_{j=0}^k\binom{k}{j}(\lambda_1\cos^2\beta)^j(\lambda_2\sin^2\beta)^{k-j},
\label{qk}
\ee
and
\be
&&(1-q(P))^{n-k}\nonumber\\
&&=\sum_{l=0}^{n-k}\binom{n-k}{l}(\lambda_1\sin^2\beta)^{n-k-l}(\lambda_2\cos^2\beta)^{l}.
\label{qnk}
\ee
Substituting Eq.~(\ref{qk}) and Eq.~(\ref{qnk}) to Eq.~(\ref{jnk}), we obtain
\be
J_{n,k}
&=&2\sum_{j=0}^k\sum_{l=0}^{n-k}\binom{k}{j}\binom{n-k}{l}\lambda_1^{n-k+j-l}\lambda_2^{k-j+l}\nonumber\\
&&\times\int_0^{\frac{\pi}{2}}d\beta(\sin\beta)^{2(n-j-l)+1}(\cos\beta)^{2(j+l)+1}.
\ee
From Eq.~(\ref{defBeta1}), we find
\be
2\int_0^{\frac{\pi}{2}}d\beta(\sin\beta)^{2(n-j-l)+1}(\cos\beta)^{2(j+l)+1}\nonumber\\
=B(n-j-l+1, j+l+1).
\ee
Furthermore, direct calculation shows
\be
&&\binom{k}{j}\binom{n-k}{l}B(n-j-l+1, j+l+1)\nonumber\\
&&=\binom{j+l}{j}\binom{n-j-l}{k-j}B(n-k+1, k+1).
\ee
Therefore, we may write
\be
J_{n,k}
&=&B(n-k+1, k+1)\nonumber\\
&&\times\sum_{j=0}^k\sum_{l=0}^{n-k}\binom{j+l}{j}\binom{n-j-l}{k-j}\lambda_1^{n-r}\lambda_2^{r},
\label{jnk2}
\ee
where 
$
r=k-j+l.
$ 
Plugging Eq.~(\ref{jnk2}) into Eq.~(\ref{target}), and comparing the obtained expression with Eq.~(\ref{mixed}), we find
\be
I_{n,k}&=&\sum_{j=0}^k\sum_{l=0}^{n-k}\binom{j+l}{j}\binom{n-j-l}{k-j}K,
\label{ink_m}
\ee
where
\be
K=\int_0^1d\lambda_1\int_0^1d\lambda_2F(\lambda_1,\lambda_2)\lambda_1^{n-r}\lambda_2^{r}.
\label{defK}
\ee
Substituting Eq.~(\ref{Fflat}) to the RHS of Eq.~(\ref{defK}) and performing the integral, we find
\be
\!\!\!\!\!\!\!\!\!\!\!\! K&&=\int_0^1d\lambda_1\lambda_1^{n-r}(1-\lambda_1)^{r}
=B(n-r+1, r+1).
\label{flat_lambda}
\ee
Plugging Eq.~(\ref{flat_lambda}) into Eq.~(\ref{ink_m}), we find Eq.~(\ref{ink_flat}).

\section{Derivation of Eq.~(\ref{ink})}
\label{app_Bures}
Similarly to Appendix \ref{app}, we first integrate $\alpha, \beta, \gamma$ and obtain Eq.~(\ref{ink_m}).
By using Eq.~(\ref{defBeta2}), the integration of $\lambda_1$ and $\lambda_2$ with the measure (\ref{FBures}) leads to a sum of the Beta functions:

\begin{widetext}
\begin{align}
K&=\frac{2}{\pi}\int_0^1d\lambda_1\Big\{\[\lambda_1^2-2\lambda_1(1-\lambda_1)+(1-\lambda_1)^2\]
\lambda_1^{n-r-\frac{1}{2}}(1-\lambda_1)^{r-\frac{1}{2}}\Big\}\nonumber\\
&=\frac{2}{\pi}\[\int_0^1d\lambda_1\lambda_1^{n-r+\frac{3}{2}}(1-\lambda_1)^{r-\frac{1}{2}}
-2\int_0^1d\lambda_1\lambda_1^{n-r+\frac{1}{2}}(1-\lambda_1)^{r+\frac{1}{2}}
+\int_0^1d\lambda_1\lambda_1^{n-r-\frac{1}{2}}(1-\lambda_1)^{r+\frac{3}{2}}\]\nonumber\\
&=\frac{2}{\pi}\[B(n-r+\frac{5}{2}, r+\frac{1}{2})-2B(n-r+\frac{3}{2}, r+\frac{3}{2})+B(n-r+\frac{1}{2}, r+\frac{5}{2})\].
\label{intF}
\end{align}
\end{widetext}
Using $yB(x+1,y)=xB(x,y+1)$, we further obtain
\be
K&=&\frac{2}{\pi}\(\frac{n-r+\frac{3}{2}}{r+\frac{1}{2}}-2+\frac{r+\frac{3}{2}}{n-r+\frac{1}{2}}\)B(n-r+\frac{3}{2}, r+\frac{3}{2})\nonumber\\
&=&\frac{2}{\pi}\frac{(n-2r)^2+n+1}{(r+\frac{1}{2})(n-r+\frac{1}{2})}B(n-r+\frac{3}{2}, r+\frac{3}{2}).
\ee
Substitution of this expression into Eq.~(\ref{ink_m}) leads to Eq.~(\ref{ink}).

\bibliography{refdeFinetti}
\end{document}